\documentclass{emulateapj}

\shorttitle{Nebular emission in high-redshift galaxies}
\shortauthors{Zackrisson et al.}

\begin{document}

\title{The impact of nebular emission on the broadband fluxes of high-redshift galaxies}
\author{E. Zackrisson\altaffilmark{1,2,3}$^*$, N. Bergvall\altaffilmark{3} \& E. Leitet\altaffilmark{3}}
\altaffiltext{*}{E-mail: ez@astro.uu.se}
\altaffiltext{1}{Tuorla Observatory, University of Turku, V\"ais\"al\"antie 20, FI-21500 Piikki\"o, Finland}
\altaffiltext{2}{Stockholm Observatory, AlbaNova University Center, 106 91 Stockholm, Sweden}
\altaffiltext{3}{Department of Astronomy and Space Physics, Box 515, 751 20 Uppsala, Sweden}

\begin{abstract}
A substantial fraction of the light emitted from young or star-forming galaxies at ultraviolet to near-infrared wavelengths comes from the ionized interstellar medium in the form of emission lines and a nebular continuum. At high redshifts, star formation rates are on average higher and stellar populations younger than in the local Universe. Both of these effects act to boost the impact of nebular emission on the overall spectrum of galaxies. Even so, the broadband fluxes and colours of high-redshift galaxies are routinely analyzed under the assumption that the light observed originates directly from stars. Here, we assess the impact of nebular emission on broadband fluxes in Johnson/Cousins $BVRIJHK$, Sloan Digital Sky Survey $griz$ and Spitzer IRAC/MIPS filters as a function of observed redshift (up to $z=15$) for galaxies with different star formation histories. We find that nebular emission may account for a non-negligible fraction of the light received from high-redshift galaxies. The ages and masses inferred for such objects through the use of spectral evolutionary models that omit the nebular contribution are therefore likely to contain systematic errors. We argue that a careful treatment of the nebular component will be essential for the interpretation of the rest-frame ultraviolet-to-infrared properties of the first galaxies formed, like the ones expected to be detected with the James Webb Space Telescope.
\end{abstract}



\keywords{Galaxies: high-redshift -- galaxies: ISM -- techniques: photometric -- radiation mechanisms: general}


\section{Introduction}  
In young galaxies, or galaxies with active star formation, massive stars photoionize the surrounding interstellar medium, thereby adding emission lines and nebular continuum flux to the integrated spectral energy distribution (SED). Since this nebular component can make up a substantial fraction of the overall light observed at rest-frame ultraviolet to near-infrared wavelengths, modellers have for a long time stressed the importance of considering the contribution from ionized gas to the SED when attempting to interpret the broadband fluxes and colours of such galaxies \citep[e.g.][]{Guiderdoni & Rocca-Volmerange a,Krüger et al.,Fioc & Rocca-Volmerange,Zackrisson et al.,Anders & Fritze-Alvensleben a}. This contribution is expected to grow with the redshift of the galaxies studied, since stellar populations are on average younger and the star formation activity more pronounced at high redshifts than in the local Universe. Even so, observers often attempt to derive ages and masses of high-redshift objects from optical to near-infrared broadband photometry data using spectral evolutionary models that only predict the SED of the stars themselves. Recent examples include studies of star-forming or young galaxies at $z\approx 2$ \citep{Shapley et al. a,Erb et al.}, $z\approx 3$ \citep{Nilsson et al.,Gawiser et al. b}, $z\approx 5$ \citep{Verma et al.}, and $z\approx 6$ \citep{Eyles et al.}. Here, we assess the reliability of this approach by estimating the fraction of nebular-to-stellar light emitted for galaxies with redshifts in the range $z=0$--15, under various assumptions about the prior star formation history.

\section{Modelling high-redshift galaxies}
We model the SEDs of galaxies using the \citet{Zackrisson et al.} spectral synthesis code, which treats nebular emission by feeding the integrated stellar component spectrum for different ages into the photoionization model Cloudy, version 90.05 \citep{Ferland et al.}. For simplicity, we assume that the nebula is spherical and radiation bounded, has a covering factor of unity, and has a gaseous metallicity $Z=0.020$ identical to that of the stars. 

Three different types of model galaxies are considered -- early-type, late-type and starburst galaxies -- differentiated by their assumed ages and star formation histories. For the early- and late-type objects, we adopt the functional form suggested by \citet{Nagamine et al.} for the temporal evolution of the star formation rate: SFR$(t)\propto (t/\tau)\exp (-t/\tau)$. While Nagamine et al. find that $\tau=1.5$ Gyr and $\tau=4.5$ Gyr give reasonable fits to the blue and red peaks in the colour distribution of low-redshift galaxies in the Sloan Digital Sky Survey (SDSS), we here adopt $\tau=0.5$ Gyr and $\tau=4.5$ Gyr for our early and late-type models, i.e. a somewhat less prolonged star formation history for early-type galaxies. This pushes the bulk of star formation ($90\%$ of the stellar mass formed) to $z>3$ (as opposed to $z>1$ for a $\tau=1.5$ Gyr model) for the early-type galaxy, making it more representative for some of the ``red and dead'' galaxies recently discovered at $z=1.5$--3 \citep[e.g][]{Labbé et al.,Kriek et al.,McGrath et al.,Stockton et al.}. 

For starburst galaxies, we assume a fixed age of 50 Myr at all $z_\mathrm{obs}$ and a constant prior SFR$(t)$. The SEDs of the early- and late-type model galaxies are evaluated at the age of the Universe at the observed redshifts we consider here: $z_\mathrm{obs}=0$--15, in steps of $\Delta z_\mathrm{obs}=0.2$. Redshifts are converted into cosmic ages assuming a $\Lambda$CDM cosmology with $\Omega_\Lambda=0.74$, $\Omega_\mathrm{M}=0.26$ and $H_0=72$ km s$^{-1}$ Mpc$^{-1}$, in accordance with the 3 year WMAP results \citep{Spergel et al.}.   

For every $z_\mathrm{obs}$, the SEDs of these model galaxies are convolved with the filter profiles of Johnson/Cousins $BVRIJHK$, SDSS $griz$, Spitzer IRAC at 3.6, 4.5, 5.8, 8.0 $\mu$m and Spitzer MIPS at 24 $\mu$m. For each filter we calculate $f_\mathrm{neb}/f_\mathrm{stars}$, the ratio of the net flux from the nebular component (the flux emitted by the nebula minus the absorbed stellar flux, the latter being essentially zero at wavelengths higher than 912 \AA) to that from the stars themselves, as a function of $z_\mathrm{obs}$. Since we here assume the nebular emission to be caused by processed far-UV starlight, flux conservation requires that the the bolometric $f_\mathrm{neb}/f_\mathrm{stars}$ will always be lower than unity. However, when integrating over broadband filter passbands at wavelengths longward of the redshifted 912 \AA{ } Lyman limit, $f_\mathrm{neb}/f_\mathrm{stars}>1$ is  possible and does occur.

For all model galaxies, we adopt a constant metallicity of $Z=0.020$ and a Salpeter initial mass function ($\mathrm{d}N/\mathrm{d}M\propto M^{-2.35}$ throughout the mass range 0.08--120 $M_\odot$). We stress that this model only predicts the SED of the stellar component and its associated photoionized nebula, and that neither dust extinction nor absorption in the gaseous medium along the line of sight are taken into account. The impact that such effects may have on the predicted $f_\mathrm{neb}/f_\mathrm{stars}$ ratios is discussed in Section 4.

\section{Results}
\begin{figure}[t]
\plotone{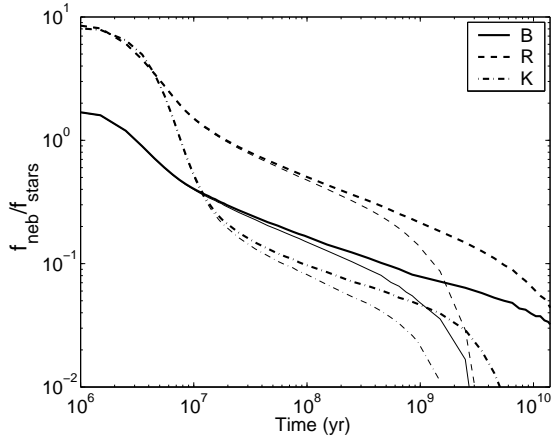}
\caption{The ratio of nebular to stellar light as a function of age in filters $B$ (solid), $R$ (dashed) and $K$ (dash-dotted) for the early- (thin lines) and late-type (thick lines) model galaxies at $z_\mathrm{obs}=0$.}
\label{fig1}
\end{figure}
To illustrate the time dependence of the nebular contribution, we plot $f_\mathrm{neb}/f_\mathrm{stars}$ in filters $BRK$ for our early- and late-type model galaxies as a function of age in Fig.~\ref{fig1}. The redshift is assumed to be $z_\mathrm{obs}=0$. Nebular emission is seen to dominate over the stellar component (i.e. $f_\mathrm{neb}/f_\mathrm{stars}>1$) at very low ages ($\sim10^6$--$10^7$ yr), and to remain non-negligible ($f_\mathrm{neb}/f_\mathrm{stars}>0.1$) up to ages of $\sim 10^8$--$10^9$ yrs, depending on the filter and star formation history. When $f_\mathrm{neb}/f_\mathrm{stars}$ is plotted against time in this fashion, without considering redshift effects, $f_\mathrm{neb}/f_\mathrm{stars}$ decreases monotonically in each filter as the stellar populations grow older. The situation becomes far more complicated if $f_\mathrm{neb}/f_\mathrm{stars}$ is instead plotted against the observed redshift of these model galaxies.
\begin{figure*}[t]
\plotone{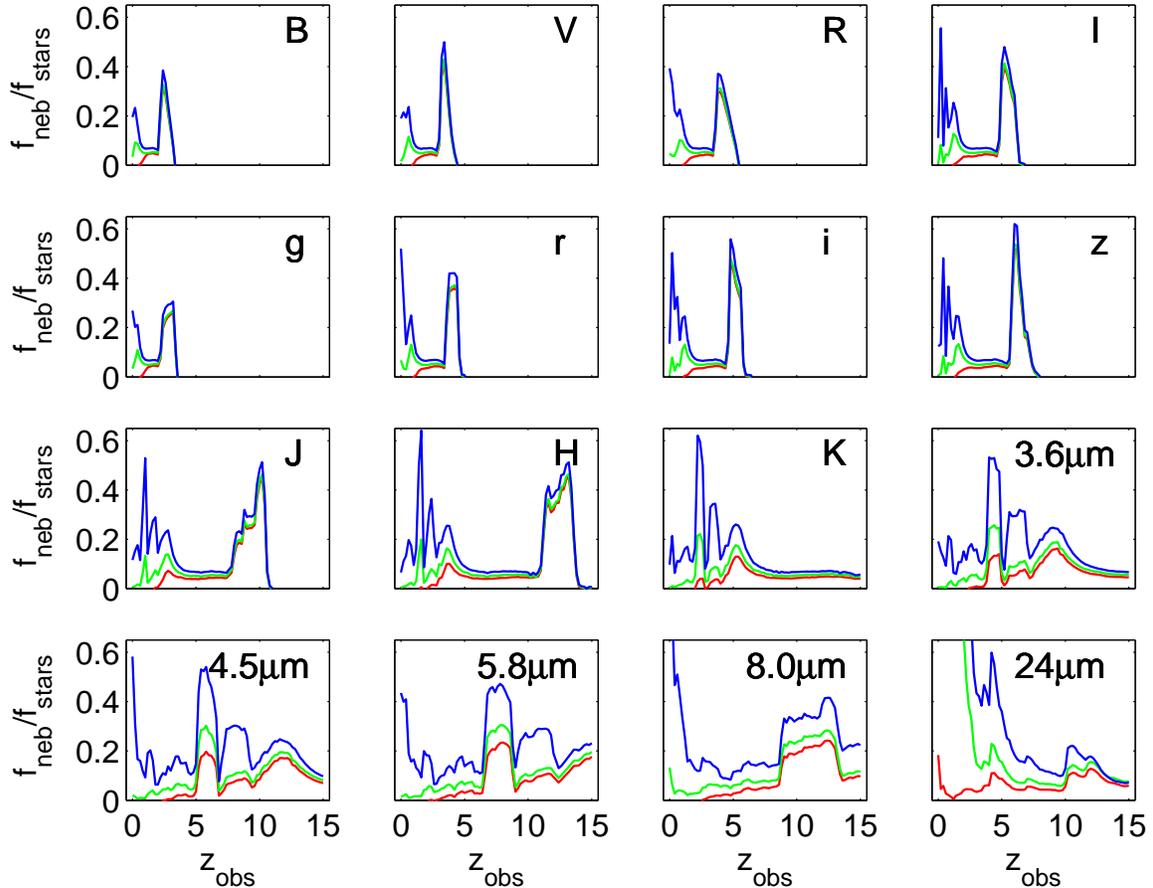}
\caption{The ratio of nebular to stellar light as a function of $z_\mathrm{obs}$ for 16 different broadband filters. The lines correspond to the different model galaxies considered: starburst galaxy (blue), late-type galaxy (green), and early-type galaxy (red). The various peaks in $f_\mathrm{neb}/f_\mathrm{stars}$ are mainly due to strong emission lines which shift in and out of the filters as a function of redshift. High values of $f_\mathrm{neb}/f_\mathrm{stars}$ signify redshift intervals where treating the observed broadband fluxes as direct starlight may be a bad approximation (see main text for details).}
\label{fig2}
\end{figure*}

In Fig.~\ref{fig2}, we plot $f_\mathrm{neb}/f_\mathrm{stars}$ as a function of $z_\mathrm{obs}$ for the three model galaxies and all the 16 filters considered. Contrary to Fig.~\ref{fig1}, $f_\mathrm{neb}/f_\mathrm{stars}$ here displays a number of maxima, most clearly seen in the case of the starburst model (blue line). 

What is the origin of these peaks in the $f_\mathrm{neb}/f_\mathrm{stars}$ ratio? For filters with central wavelengths in the range from $B$ to $H$, two sets of  peaks at low and high $z_\mathrm{obs}$ are clearly visible (e.g. at $z_\mathrm{obs}\approx0$--3 and $z_\mathrm{obs}\approx8$--11 in the $J$ band), with a valley of low $f_\mathrm{neb}/f_\mathrm{stars}$ values inbetween. The wide peak seen at high $z_\mathrm{obs}$ in filters with central wavelengths in the range $B$--$H$ is caused by the Ly$\alpha$ emission line. The peaks at low $z_\mathrm{obs}$ come from a number of different spectral features. In filters $B$ and $g$, these peaks are mainly due to the H$\beta$ emission line and nebular Balmer continuum. These emission processes are accompanied by the [O III]$\lambda\lambda$4959,5007 emission lines in filters $VRI$ and $riz$, and by the H$\alpha$ line in $RI$ and $riz$. The $f_\mathrm{neb}/f_\mathrm{stars}$ peaks due to Balmer continuum, H$\beta$, [O III] and H$\alpha$ become broader as they redshift into the wider filters at longer wavelengths. The most prominent of the bumps seen at $z_\mathrm{obs}>2$ for filters with central wavelengths in the range from $K$ to 8.0 $\mu$m is due to H$\alpha$.  

For each filter, $f_\mathrm{neb}/f_\mathrm{stars}$ is only plotted up to the redshift where the Lyman limit at 912 \AA{ }enters the filter. When this happens, absorption within the nebula begins to dominate over emission processes, and galaxies become increasingly more difficult to detect because of the very low stellar flux emerging shortward of this limit. For computational reasons, absorption shortward of the Lyman limit is here treated as a negative flux, which means that $f_\mathrm{neb}$ becomes negative and drops below the plotted range at this redshift. This prevents the $f_\mathrm{neb}/f_\mathrm{stars}$ curves to extend all the way up to $z_\mathrm{obs}=15$ for filters in the wavelength range from $B$ to $J$.

The rise in $f_\mathrm{neb}/f_\mathrm{stars}$ seen at low $z_\mathrm{obs}$ in filters at 4.5--24 $\mu$m is partly due to a number of strong emission lines, but in the 8 and 24 $\mu$m filters, there is also a considerable contribution from nebular continuum at $z_\mathrm{obs}\approx 0$. Even though the flux from the nebula can become much higher than the flux of the stars at the latter two wavelengths, this is in fact of little relevance for observations, since the overall SED is likely to be dominated by dust emission in these filters at low $z_\mathrm{obs}$, and not by stars or nebular emission.

Even if we disregard the unobservable behaviour of $f_\mathrm{neb}/f_\mathrm{stars}$ at 8--24 $\mu$m at $z_\mathrm{obs}\approx 0$, certain combinations of filters and redshifts produce $f_\mathrm{neb}/f_\mathrm{stars}$ of up to $\approx 60$\% for all the model galaxies considered. This means that estimates of the stellar population mass, based on the incorrect assumption that all of the observed flux is directly produced by stars, may be overestimated by this fraction. Moreover, $f_\mathrm{neb}/f_\mathrm{stars}$ changes dramatically from one filter to the next for fixed $z_\mathrm{obs}$, which implies that nebular emission will alter the colours of the overall SED. As an example, consider the difference in $f_\mathrm{neb}/f_\mathrm{stars}$ between the $J$ and 4.5 $\mu$m filters at $z_\mathrm{obs}=6$ for the starburst model. Nebular emission will change this colour by $\approx 0.4$ mag, which means that the best-fitting age is likely to be substantially affected. While a first-order correction for nebular emission can be applied to observed broadband fluxes by assessing the likely equivalent widths of the most prominent emission lines (e.g. Ly$\alpha$, H$\beta$, [O III], H$\alpha$), high-accuracy interpretations of broadband fluxes will require a consideration of the entire nebular SED, at least in the case of starburst-type galaxies. 

\section{Discussion}
Our estimates indicate that nebular emission may be an important, and hitherto largely neglected ingredient in models used to extract ages and masses from broadband fluxes of high-redshift galaxies. We note that its relative contribution is comparable to that of recent updates of the evolution of thermally pulsating asymptotic giant branch stars \citep[e.g.][]{Maraston,Maraston et al.,Bruzual}, which have recently attracted a lot of attention in the high-redshift community. A careful treatment of the nebular component is therefore likely to be very important for the interpretation of the ultraviolet-to-infrared properties of the first galaxies formed, like the ones expected to be detected with the James Webb Space Telescope\footnote{\url{http://www.jwst.nasa.gov/}}.

However, the $f_\mathrm{neb}/f_\mathrm{stars}$ ratios presented here represent no more than tentative estimates, since there are a number of mechanisms that could boost or diminish the effects of nebular emission, and which are difficult to control at the current time.  

\begin{figure}[t]
\plotone{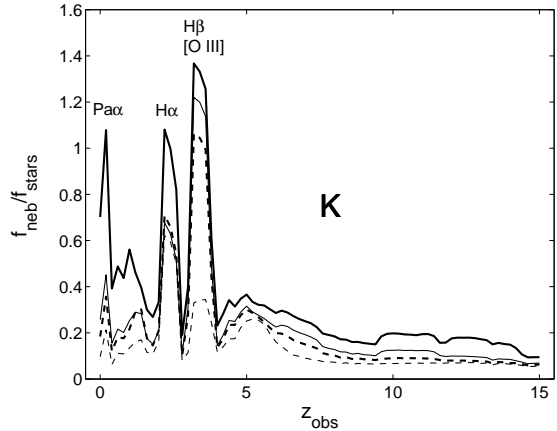}
\caption{The ratio of nebular to stellar light as a function of $z_\mathrm{obs}$ in $K$ for our starburst galaxy model at $Z=0.001$ (thick solid), 0.004 (thin solid) 0.008 (thick dashed) and 0.020 (thin dashed). The latter metallicity corresponds to that plotted in Fig.~\ref{fig2}. The emission lines responsible for the most prominent features have been indicated by labels.}
\label{fig3}
\end{figure}

Owing to the likely breakdown of the closed-box approximation at high redshifts, we have here adopted a fixed stellar and gaseous  metallicity of $Z=0.020$ instead of implementing an age-dependent metallicity. While appropriate for many types of galaxies at $z_\mathrm{obs}\approx 0$, the typical metallicity must realistically be much lower at earlier epochs, even though the exact redshift evolution is difficult to assess. The choice of $Z=0.020$ ensures that our $f_\mathrm{neb}/f_\mathrm{stars}$ ratios are conservative, since lower metallicities would imply higher Lyman continuum fluxes from the stars and less efficient metal cooling of the gas, resulting in boosted $f_\mathrm{neb}/f_\mathrm{stars}$ ratios. This is demonstrated in Fig.~\ref{fig3}, where we have plotted the redshift evolution of $f_\mathrm{neb}/f_\mathrm{stars}$ in the $K$ band for our starburst model galaxy at various lower metallicities. Lowering $Z$ increases the relative nebular contribution at essentally all $z_\mathrm{obs}$, even giving $f_\mathrm{neb}/f_\mathrm{stars}\gtrsim 1$ in three different redshift intervals. The emission lines responsible for the most prominent $f_\mathrm{neb}/f_\mathrm{stars}$ peaks in this passband have been indicated with labels. Since $f_\mathrm{neb}$ and $f_\mathrm{stars}$ respond to metallicity changes in different ways, the exact behaviour of $f_\mathrm{neb}/f_\mathrm{stars}$ as a function of $Z$ and $z_\mathrm{obs}$ actually turns out to be quite complex. As an example, consider the H$\alpha$-dominated peak at $z_\mathrm{obs}\approx 2.2$, which shows very little metallicity dependence in the range from $Z=0.020$ to $Z=0.004$, but increases rapidly from $Z=0.004$ to $Z=0.001$. The H$\alpha$ luminosity changes relatively little between these metallicities, and the sudden increase in $f_\mathrm{neb}/f_\mathrm{stars}$ at $Z=0.001$ is instead caused by a drop in $f_\mathrm{stars}$ related to properties of the adopted stellar evolutionary tracks. By contrast, the rapid rise of $f_\mathrm{neb}/f_\mathrm{stars}$ at $z_\mathrm{obs}\approx 3.2$ between $Z=0.020$ and $Z=0.008$, followed by a less dramatic growth at lower $Z$, is caused by a combination of the nonmonotonic metallicity dependence of the [O III]/H$\beta$ emission-line ratio  \citep[see Fig. 17 in ][]{Nagao et al.} and the drop in $f_\mathrm{stars}$ at $Z=0.001$.

We have also adopted the conservative assumption of a Salpeter initial mass function throughout the redshift range $z_\mathrm{obs}=0$--15, whereas theoretical arguments suggest that the initial mass function should turn top-heavy at high redshifts, possibly introducing very massive Population III stars \citep[for a review, see][]{Bromm & Larson}. Because of the high Lyman continuum fluxes of such stars, this would increase $f_\mathrm{neb}/f_\mathrm{stars}$ at high $z_\mathrm{obs}$. 

Our estimates of the $f_\mathrm{neb}/f_\mathrm{stars}$ ratios are based on the assumption that gas is ionized by stars only. The presence of active galactic nuclei would boost the flux of ionizing photons, and shock heating of the gas may also become important in galaxies undergoing mergers. Such effects are expected to increase $f_\mathrm{neb}/f_\mathrm{stars}$ in certain high-redshift objects.

Our model assumes that all Lyman continuum photons are absorbed by gas close to the galaxy and therefore contribute to the nebular SED, but this may be unrealistic, since dust extinction shortward of the Lyman limit would substantially reduce the hydrogen-ionizing flux \citep[e.g][]{Inoue}, and hence decrease $f_\mathrm{neb}/f_\mathrm{stars}$. The exact level of Lyman continuum extinction is difficult to assess even in the local Universe, and even more so at higher redshifts. This treatment also implicitly assumes that no Lyman continuum photons can escape into the intergalactic medium, either through holes in the gas (i.e. a covering factor below unity) or through a matter-bounded nebula (i.e. insufficient gas supply to form a complete Str\"omgren sphere). This approximation must of course break down if galaxies are to contribute to cosmic reionization, and a few detections of Lyman continuum leakage have indeed been claimed at both low and high redshifts \citep[e.g.][]{Steidel et al.,Bergvall et al.,Shapley et al. b}. Any mechanism that allows Lyman continuum photons to escape would lead to a nebular flux lower than that predicted here.

Since Ly$\alpha$ photons are resonantly scattered by neutral hydrogen both inside the source galaxies and in the intergalactic medium, the strength of the Ly$\alpha$ line becomes very sensitive to the assumed geometry and kinematical properties of neutral hydrogen along the line of sight. Since such effects are not taken into account here, the Ly$\alpha$ contribution to $f_\mathrm{neb}/f_\mathrm{stars}$ that is relevant for observers is likely to be lower than that predicted by us. This affects the height of the peak at high $z_\mathrm{obs}$ in the $f_\mathrm{neb}/f_\mathrm{stars}$ evolution for filters with central wavelengths in the range $B$--$H$ in Fig.~\ref{fig2}. At the current time, estimates of the Ly$\alpha$ escape fraction $f_\mathrm{esc,Ly\alpha}$ vary greatly. A Ly$\alpha$ escape fraction as low as $f_\mathrm{esc,Ly\alpha}\approx 0.02$ \citep[e.g.][]{Le Delliou et al.} would make the impact of Ly$\alpha$ on $f_\mathrm{neb}/f_\mathrm{stars}$ completely insignificant, whereas $f_\mathrm{esc,Ly\alpha}\approx 0.8$ \citep[e.g.][]{Gawiser et al. a,Kobayashi et al.} would imply just a slightly smaller effect than that indicated in Fig.~\ref{fig2}. Either way, the increased Ly$\alpha$ opacity at epochs prior to complete cosmic reionization \citep[$z_\mathrm{obs}\gtrsim 6$; see e.g.][for a review]{Fan et al.} should make the Ly$\alpha$ flux smaller than assumed here. Hence, the $f_\mathrm{neb}/f_\mathrm{stars}$ peaks seen at redshifts higher than this in panels $J$ and $H$ in Fig.~\ref{fig2} are likely to be too prominent.

Whereas a uniform, foreground, screenlike distribution of dust would decrease the nebular and stellar fluxes by an equal factor, thereby leaving $f_\mathrm{neb}/f_\mathrm{stars}$ unchanged, selective extinction, i.e. the fact that the young stellar population and the associated ionized gas may have different dust properties than the older stars, could in principle drive $f_\mathrm{neb}/f_\mathrm{stars}$ either way. Observations of starburst galaxies suggest that the nebular component may suffer more extinction than the old stars \citep[e.g.][]{Calzetti et al.}, but the opposite effect has also been advocated to explain the equivalent widths of Ly$\alpha$ in high-redshift galaxies \citep{Finkelstein et al.}.
 
\acknowledgments{E.Z. acknowledges research grants from the Swedish Research Council, the Royal Swedish Academy of Sciences and the Academy of Finland. We wish to thank the anonymous referee for useful comments which helped improve the quality of this Letter.}

\end{document}